\documentclass[twocolumn,superscriptaddress,aps,prb]{revtex4-2}

\usepackage{graphicx}
\usepackage{dcolumn}
\usepackage{bm}
\usepackage{subfigure}
\usepackage{amsmath}
\usepackage{mathrsfs}
\usepackage{amssymb}
\usepackage{setspace}
\usepackage{array}
\usepackage{multirow}
\usepackage{float}
\usepackage{flushend}
\usepackage{footmisc}

\usepackage[pdfstartview=FitH,
CJKbookmarks=true,
colorlinks,
linkcolor=blue,
anchorcolor=blue,
citecolor=blue,
urlcolor=blue,
]{hyperref}

\usepackage{footmisc}

\renewcommand{\footnoterule}{
	\kern -4pt  
	\hrule width 0.18\linewidth height 0.6pt
	\kern 12pt 
}

\usepackage{footmisc}

\usepackage{braket}

\begin{document}
	
	\preprint{APS/123-QED}
	
	\title{Dissipation induced transition between delocalization and localization in the three-dimensional Anderson model}
	
	\author{Xuanpu Yang}
	\affiliation{School of Physics, Nankai University, Tianjin 300071, China}
	
	\author{Xiang-Ping Jiang}
	\affiliation{Zhejiang Lab, Hangzhou 311121, China}
	
	\thanks{X Yang and X-P Jiang contributed equally to this work.}

	\author{Zijun Wei}
	\affiliation{School of Physics, Nankai University, Tianjin 300071, China}
	
	\author{Yucheng Wang}
	\email{wangyc3@sustech.edu.cn}
	\affiliation{Shenzhen Institute for Quantum Science and Engineering,
		Southern University of Science and Technology, Shenzhen 518055, China}
	\affiliation{International Quantum Academy, Shenzhen 518048, China}
	\affiliation{Guangdong Provincial Key Laboratory of Quantum Science and Engineering, Southern University of Science and Technology, Shenzhen 518055, China}

	\author{Lei Pan}%
	\email{panlei@nankai.edu.cn}
	\affiliation{School of Physics, Nankai University, Tianjin 300071, China}

	\begin{abstract} 
		
		We investigate the probable delocalization-localization transition in open quantum systems with disorder. The disorder can induce localization in isolated quantum systems and it is generally recognized that localization is fragile under the action of dissipations from the external environment due to its interfering nature. Recent work [Y. Liu, et al, \href{https://journals.aps.org/prl/abstract/10.1103/PhysRevLett.132.216301}{Phys. Rev. Lett. {132}, 216301 (2024)}] found that a one-dimensional quasiperiodic system can be driven into the localization phase by a tailored local dissipation where a dissipation-induced delocalized-localized transition is proposed. Based on this, we consider a more realistic system and show that a dissipation-induced transition between delocalization and localization appears in the three-dimensional (3D) Anderson model. By tuning local dissipative operators acting on nearest neighboring sites, we find that the system can relax to localized states dominated steady state instead of the choice of initial conditions and dissipation strengths. Moreover, we can also realize a delocalized states predominated steady-state from a localized initial state by using a kind of dissipation operators acting on next nearest neighboring sites. 
		Our results enrich the applicability of dissipation-induced localization and identify the transition between delocalized and localized phases in 3D disordered systems.

	\end{abstract}
	
	
	\maketitle
	\section{Introduction}
	Disorder plays a vital role in condensed matter physics, permeating nearly all its subfields. One of the most significant physical consequences of disorder is the phenomenon of Anderson localization \cite{Anderson1958}. The three-dimensional (3D) Anderson model serves as a paradigmatic framework for studying the Anderson transition, where a localization transition occurs in a lattice when the disorder strength exceeds a critical threshold. Near this transition point, mobility edges (MEs) emerge, marking the energy boundaries that separate localized states from delocalized states. These MEs are a hallmark feature of the 3D Anderson model and have profound implications for understanding transport properties, such as electronic conductivity and heat conduction in disordered systems.
	The existence of MEs depends on the spatial dimension and the disorder type \cite{AndersonTrans2008,Anderson_Scalling}. For example,  MEs are absent in one- and two-dimensional random disordered systems, where all wave functions are localized with arbitrary disorder strength. But MEs may occur in some one-dimensional (1D) systems with the quasiperiodic potential  \cite{1D_ME1,1D_ME2,1D_ME3,1D_ME4,1D_ME5,1D_ME6,1D_ME7,1D_ME8,1D_ME9,1D_ME10,1D_ME11,1D_ME12}. 
	
	Realistically, a system always inevitably interacts with external environments. The coupling between the system and degrees of freedom of the environment leads to dissipation, which makes the system relax to a specific steady state. 
	A quantum system with a coupled reservoir constitutes an open quantum system whose time-evolution is determined by non-unitary dissipative dynamics rarher than the unitary dynamics of closed systems. The research of open quantum systems has a long history of research and has renewed interest and received fast-growing attention in recent years, fueled by experimental advancements in precisely controlling dissipation and system parameters \cite{Exp1,Exp2,Exp3,Exp4,Exp5,Exp6,Exp7,Exp8}. 
	Theoretical research on open quantum systems has also made significant strides, revealing that the interplay between dissipation and inter-particle interactions can lead to a wealth of novel phenomena \cite{OpenMB1,OpenMB2,OpenMB3,OpenMB4,OpenMB5,OpenMB6,OpenMB7,OpenMB8,OpenMB9,OpenMB10,OpenMB11,OpenMB12,OpenMB13,OpenMB14,OpenMB15,OpenMB16,OpenMB17,OpenMB18,OpenMB19,OpenMB20,OpenMB21,OpenMB22,OpenMB23,OpenMB25,OpenMB26}. 
	Moreover, open quantum systems with disordered or quasiperiodic potential have also attracted widespread attention  \cite{OpenMBL1,OpenMBL2,OpenMBL3,OpenMBL4,OpenMBL5,OpenDisorder1,OpenDisorder2,OpenDisorder3,OpenDisorder4,OpenDisorder5,OpenDisorder6,OpenDisorder7,OpenDisorder8,OpenDisorder9,OpenDisorder10,OpenDisorder11,OpenDisorder12,OpenDisorder13,OpenDisorder14,OpenDisorder15,OpenDisorder16,OpenDisorder17,OpenDisorder18}. 
	The investigation of disordered and dissipative systems, particularly within the framework of the 3D Anderson model, has attracted considerable attention owing to its profound implications. Recent research efforts have been directed towards elucidating the localization-delocalization transition in such systems, with a specific focus on the influence of dissipation. For instance, a study in Ref.\cite{OpenDisorder17} explored the statistical properties of eigenvalue distributions for non-Hermitian symmetric random matrices as they approach the Anderson localization transition and Ref.\cite{OpenDisorder18} has provided critical insights into how dissipation facilitates transitions from localized to delocalized states.
	
	Previous studies on the interplay between dissipation and transport have demonstrated that environment-induced dephasing can disrupt localization and enhance transport \cite{OpenDisorder7,OpenDisorder8,OpenDisorder9}. Consequently, it has been widely believed that Anderson localization is fragile in the presence of dissipation, as the coherence underlying localization is easily destroyed, leading to delocalized steady states. However, this traditional view has recently been challenged. For instance, Yusipov et al. \cite{Yusipov} demonstrated that Anderson localization can persist in the steady state when dissipative operators are carefully chosen. Building on this, Y. Liu et al. \cite{WYC_PRL} generalized the theory to 1D quasiperiodic systems with MEs, showing that specific forms of dissipation can drive the system into either delocalized or localized steady states. This implies that dissipation can induce transitions between delocalized and localized phases, rather than simply destroying localization. These findings hold significant theoretical and practical value for understanding open quantum systems with disorder.
	
	This paper further promotes previous significant results by exploring a more realistic model, i.e., the 3D Anderson model. The rest of this paper is organized as follows. We begin with a brief synopsis of the localization property of the 3D Anderson model in the Sec. \ref{Property}. Then, we briefly introduce the dissipative 3D Anderson model as an open system with selected dissipation operators in Sec. \ref{Model}.  
	In Sec. \ref{dissipation_localization}, we discuss the dissipation-induced localization and then the dissipation-induced delocalization in Sec. \ref{dissipation_extension}.
	And finally, we summarize in Sec. \ref{summarize}.

	\section{three-dimensional Anderson model}
	\label{Property}

\begin{figure}[!ht]
		\includegraphics[width=8.5cm]{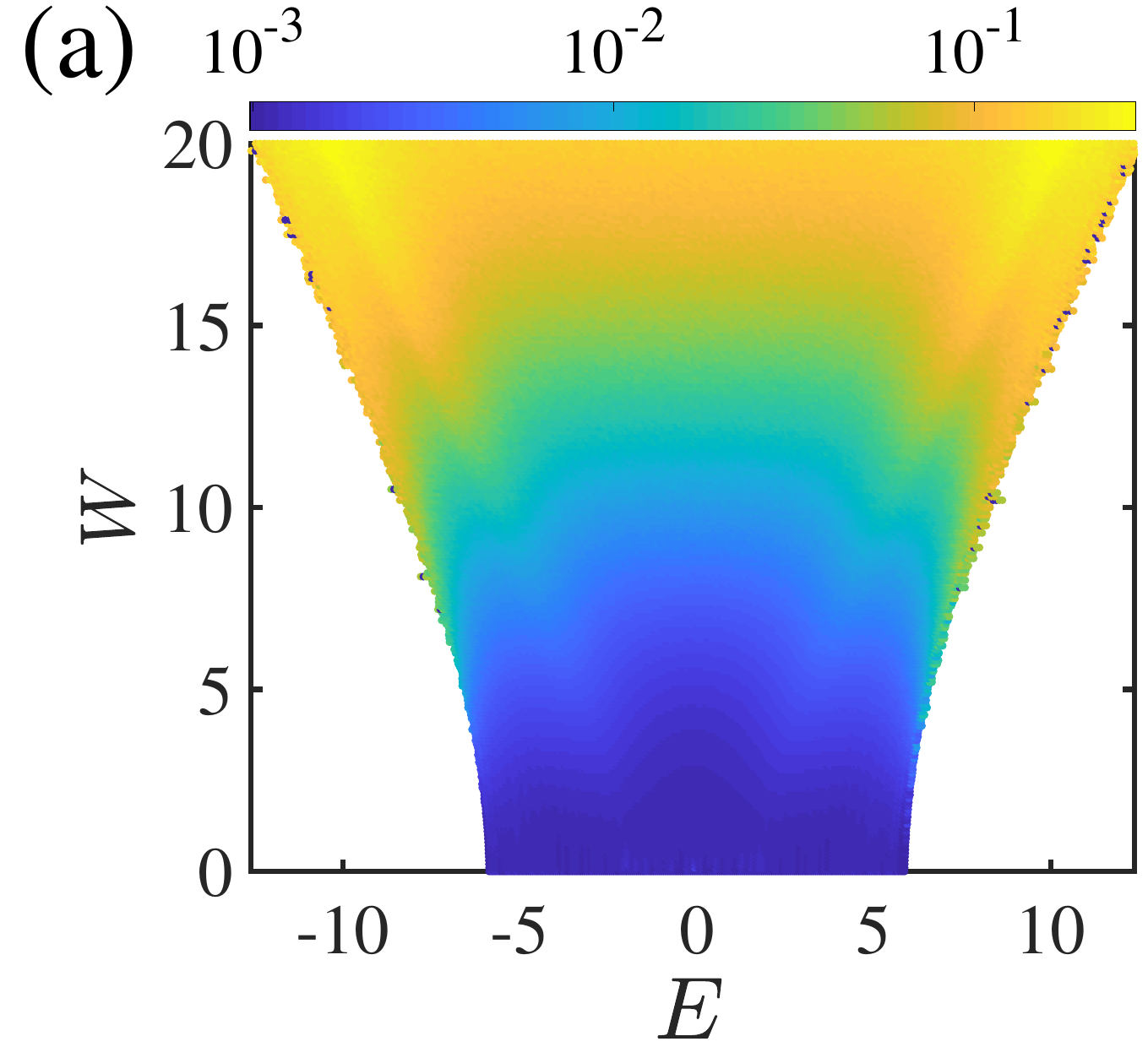}
		\includegraphics[width=8.5cm]{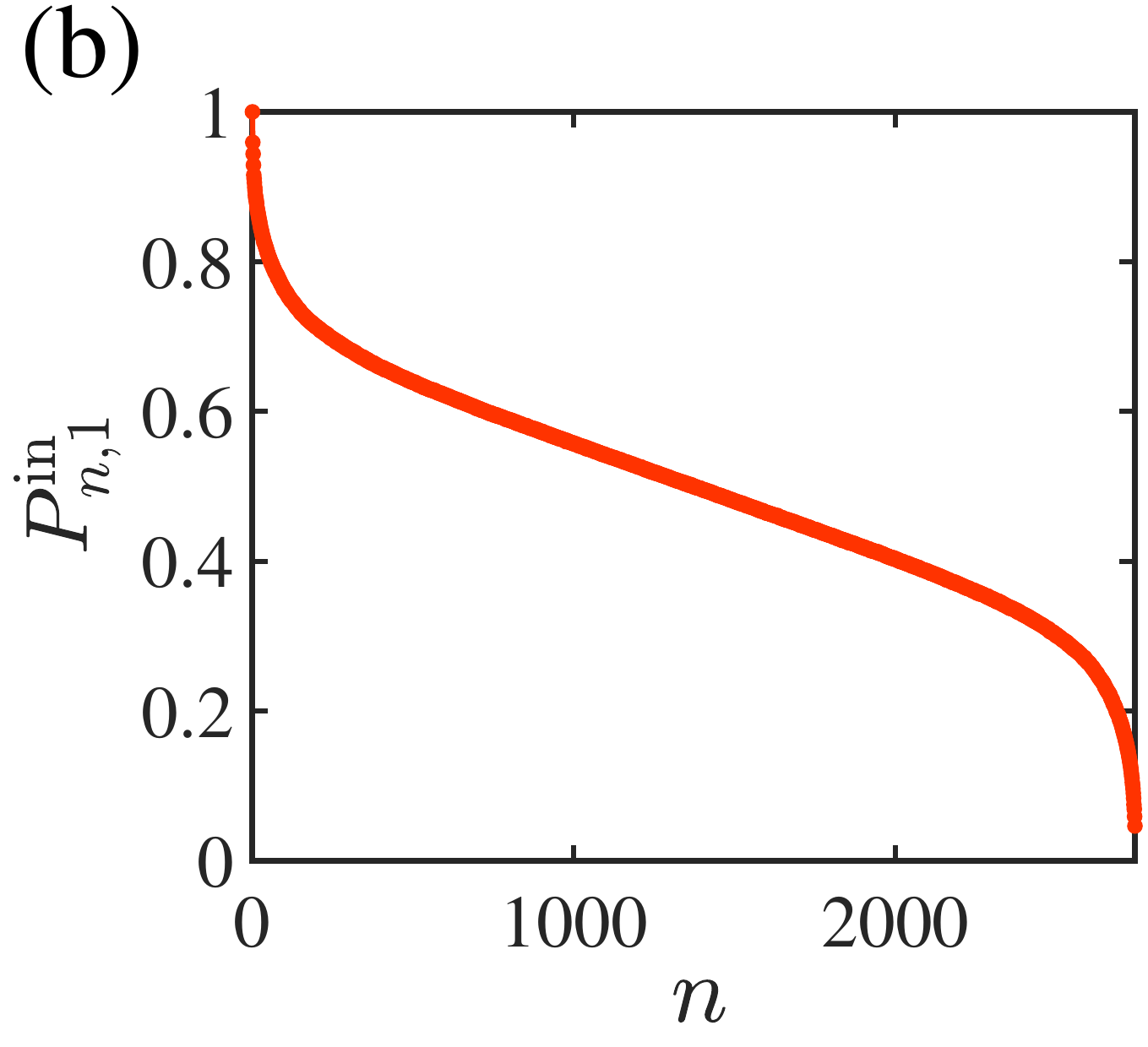}
		\caption{(a) The Inverse Participation Ratio (IPR) of the 3D Anderson model in the open boundary condition with system length $L=14$. The IPR is plotted as a function of energy eigenvalues and disorder strength, with the numerical values of IPR represented by color distribution. The figure depicts the transition between localized and delocalized states as the disorder strength varies from $0$ to $20$. When the disorder strength is small, eigenstates of the system are predominantly composed of an delocalized state. As the disorder strength increases, states with larger absolute energy eigenvalues correspond to localized states, while states with smaller absolute energy eigenvalues correspond to delocalized states.
			(b) The proportion of in-phase site pairs $P^{\text{in}}_{n,1}$ for each eigenstate where the low-energy (high-energy) state has a larger (smaller) value. Here the system length and disorder strength are taken as $L=14$ and $W=13$ respectively, and we calculate the IPR and $P^{\text{in}}_{n,1}$ over $100$ disorder average.}
		\label{fig_IPR_Phase}
	\end{figure}

	We begin by introducing the 3D Anderson model and required dissipation operators needed for this paper.
	The Hamiltonian of the 3D Anderson model is written as
	\begin{align}
		H_S = \sum_{j} \epsilon_{j} c_{j}^\dagger c_{j} + J\sum_{\langle i,j\rangle}\left( c_{i}^\dagger c_{j}+\text{H.c.}\right),
	\end{align}
	where $c_{j}^\dagger$ and $c_{j}$  are the
	creation and annihilation operators of a particle on the $j$-th
	site. The first term is the on-site term with uncorrelated random energies $\epsilon_{j}$ which are chosen as a standard uniform distribution $\epsilon_j \in \left[ -\frac{W}{2}, \frac{W}{2} \right]$ with the disorder strength $W$. The second term is the hopping term with strength $J$ where $\langle i,j\rangle$ represents nearest-neighbor sites. As the disorder strength increases, the Anderson transition occurs at a critical value $W_c\approx16.5$ \cite{AL_scalling1,AL_scalling2}. Beyond this critical value, all wave functions become exponentially localized. Near, but just below this critical value, there exist MEs where wave functions close to the center of the energy spectrum are spatially delocalized, while those at the edges of the spectrum are spatially localized. To provide an understanding of the Anderson transition, we take advantage of the IPR as a criterion to distinguish between delocalized and localized states. The IPR is defined as
	\begin{align}
		I=\frac{\sum_{\boldsymbol{r}}|\psi(\boldsymbol{r})|^4}{(\sum_{\boldsymbol{r}}|\psi(\boldsymbol{r})|^2)^2},
	\end{align}
	with the eigen-function $\psi(\boldsymbol{r})$.
	The inverse participation ratio (IPR) is a measure of how many localized bases participate in the eigenstate of the system. For delocalized states, the IPR is the order of magnitude $I \sim 1 / N_{\rm {tot}}$ with a total number of sites $N_{\rm {tot}}$ meaning the IPR tends to zero in the thermodynamics. In contrast, for localized states, the wavefunction occupies only a few sites and is independent of the system size, resulting in $I$ being a finite non-zero value even as the size approaches infinity.
	To show the Anderson transition, in Fig.\ref{fig_IPR_Phase}(a), we plot the IPR of the 3D Anderson model on a cubic lattice (with side length $L$) as a function of disorder strength $W$ and energy $E$ with disorder average. 
	One can see that for weak disorder regimes, eigenstates in the whole energy spectrum are delocalized with $I \sim 1/L^3$. As we increase the disorder strength to nearly the critical value $W_c $, the corresponding IPR becomes three orders of magnitude larger, which exhibits the transition from metal to insulator phase. Moreover,
	it also shows that localized states have already emerged in 
	both edges of the energy spectrum below the critical value $W<W_c$.
	The energy threshold $E_c$ that separates delocalized and localized states for a fixed disorder strength is referred to as
	the ME. 
	
	To further discuss properties of eigenstates, we then explore the phase difference between two lattice sites. For any eigenstate $|\psi_{n}\rangle=\sum_{\boldsymbol{r}}\psi_{n}(\boldsymbol{r})c_{\boldsymbol{r}}^{\dagger}|\text{vac}\rangle$, we compute the phase difference $\Delta\phi^n_{j,l}$ between two sites at $\boldsymbol{r}$ and $\boldsymbol{r+l}$ along each direction as  $\Delta\phi^n_{\boldsymbol{r},\boldsymbol{l}}\equiv(\Delta\phi^n_{\boldsymbol{r},\boldsymbol{l}_x},\Delta\phi^n_{\boldsymbol{r},\boldsymbol{l}_y},\Delta\phi^n_{\boldsymbol{r},\boldsymbol{l}_x})$ where $\Delta\phi^n_{\boldsymbol{r},\boldsymbol{l}_\lambda}=\arg(\psi_{n}(\boldsymbol{r}))-\arg(\psi_{n}(\boldsymbol{r}+\boldsymbol{l}_\lambda))$ with $\lambda=x,y,z$. We refer to the phase difference $\Delta\phi^n_{\boldsymbol{r},\boldsymbol{l}_\lambda}=0$ ($\Delta\phi^n_{\boldsymbol{r},\boldsymbol{l}_\lambda}=\pi$) as in-phase (out-of-phase) and meanwhile calculate the number of in-phase pairs $N_{n,l}$ with distance $l$.  Then we obtain the proportion of in-phase pairs $P^{\text{in}}_{n,l}=N_{n,l}^{\text{in}}/N_t$ where $N_t$ denotes the total number of pairing sites along three axes. Fig.\ref{fig_IPR_Phase}(b) plots $P^{\text{in}}_{n,l}$ of eigenstates for $l=1$ which illustrates that the higher (lower) energy has smaller (larger) $P_{n,1}^{\text{in}}$ for $W<W_c$. 
	Such consequence is related to the localization property of stationary states.

	\section{Dissipative three dimensional Anderson model}
	\label{Model}
	To investigate behaviors of the 3D Anderson model under dissipation, we consider the system couples a reservoir whose 
	total Hamiltonian $H_T$ is given by
	\begin{align}
		H_T=H_S+H_R+H_{SR}
	\end{align}
	where $H_S, H_R$ are Hamiltonians respectively belonging to the system and reservoir, and $H_{SR}$ denotes the coupling between them. After tracing the reservoir's degrees of freedom under the Born-Markov approximation \cite{Moy1999,Breuer2002}, the dynamical evolution of the system can be expressed by the following Lindblad form \cite{Lindblad1,Lindblad2}
	
	\begin{align}
		\frac{d \rho(t)}{dt} = \mathscr{L}[\rho(t)] = -i [H_S, \rho(t)] + \mathcal{D}[\rho(t)]\label{Lindblad}
	\end{align}
	where $\rho(t)$ is the reduced density matrix and $\mathcal{L}$ is referred to as the Liouvillian superoperator. The first term in the right hand side of Eq.\eqref{Lindblad} is the coherent evolution and the second term represents dissipative evolution, which is given by
	\begin{align}
		\mathcal{D}[\rho(t)] = \sum_{j}\sum_{m=1}^{M} \Gamma_{j}^{(m)} \left( O_{j}^{(m)} \rho O_{j}^{(m)\dagger} - \frac{1}{2} \Big\{  O_{j}^{(m)\dagger} O_{j}^{(m)}, \rho \Big\} \right).\label{Dissipator}
	\end{align}
	Here $\{,\}$ denotes the anticommutator, and $O_{j}^{(m)}$ is jump operator. $j$ is the lattice site index and $M$ represents the number of dissipation channels on the site with dissipation strength $\Gamma_{j}^{(m)}$ on each channel. 
	
	Using the Choi-Jamiołkowski isomorphism \cite{CJ1,CJ2}, the Lindblad equation can be written as an equivalent form $\frac{d}{dt}|\rho\rangle=\mathscr{L}|\rho\rangle$ where $|\rho\rangle=\sum_{i, j} \rho_{i, j}|i\rangle \otimes|j\rangle$ is vectorized density matrix with matrix element $\rho_{i, j}$. In this way, the Liouvillian superoperator would be expressed as (\ref{appendix:A}) 
	\begin{align}
		\mathscr{L}= & -i\left(H \otimes I-I \otimes H_{S}^{\mathrm{T}}\right) \nonumber \\
		& +\sum_{j}\sum_{m=1}^{M}\Bigg[2 O_{j}^{(m)} \otimes O^{*(m)}_{j}-O^{(m)\dagger}_{j} O_{j}^{(m)} \otimes I\nonumber \\
		&-I \otimes\left(O^{(m)\dagger}_{j} O_{j}^{(m)}\right)^{\mathrm{T}}\Bigg]. \label{Liouvillian}
	\end{align}
	
	As with the Hamiltonian determining dynamics of a closed quantum system, for an open quantum system, the dissipative dynamics is determined by the spectrum of Liouvillian superoperator \eqref{Liouvillian} and the formal solution is 
	$|\rho(t)\rangle=e^{\mathscr{L}t}|\rho(0)\rangle$
	An open quantum system reaches its steady state $|\rho_{ss}\rangle= \lim_{t \to \infty} |\rho(t)\rangle$ being the eigenstate with zero eigenvalue of $\mathscr{L}$. The steady state is closely related to the choice of jump operators in the Liouvillian superoperator. Here we consider jump operators of the following form
	
	\begin{align}
		\begin{split}
			O_{j}^{(1)} = \left( c_{\boldsymbol{j}}^\dagger+e^{i\alpha}c_{\boldsymbol{j} + \boldsymbol{l}_x}^\dagger \right) 
			\left( c_{\boldsymbol{j}} -e^{i\alpha} c_{\boldsymbol{j} + \boldsymbol{l}_x}\right),\\
			O_{j}^{(2)} = \left( c_{\boldsymbol{j}}^\dagger+e^{i\beta}c_{\boldsymbol{j} + \boldsymbol{l}_y}^\dagger \right) 
			\left( c_{\boldsymbol{j}} -e^{i\beta} c_{\boldsymbol{j} + \boldsymbol{l}_y}\right),\\
			O_{j}^{(3)} = \left( c_{\boldsymbol{j}}^\dagger+e^{i\gamma}c_{\boldsymbol{j} + \boldsymbol{l}_z}^\dagger \right) 
			\left( c_{\boldsymbol{j}} -e^{i\gamma} c_{\boldsymbol{j} + \boldsymbol{l}_z}\right),\\
		\end{split}
		\label{O_j}
	\end{align}
	
	This type of operators first were introduced in Refs.\cite{Jump1,Jump2}. The physical implementation is based on a 1D Bose-Hubbard chain \cite{BHchain}, and another setup is proposed on optical Raman lattices for $0$ or $\pi$ phases \cite{WYC_PRL}. Physically, above each jump operator acts on a pair of sites along an axis and changes the relative phase between them. In particular, these operator tends to synchronize two sites from an in-phase (out-of-phase) mode to an out-of-phase (in-phase) one for the dissipation phase $\alpha,\beta,\gamma$ being set to $\pi$ ($0$). That's to say, for $ \alpha = \beta = \gamma = 0$, the jump operators annihilate an out-of-phase state and then create an in-phase state while for the case $\alpha = \beta = \gamma = \pi$, they annihilate an in-phase state and then create an out-of-phase state. 
		The relative phase between the jump operators acting on neighboring sites plays a crucial role in determining the system's localization properties. By appropriately choosing dissipation phases, the system can be driven into a specific region (low-energy, high-energy or middle region) in the spectrum. Such property is crucial for understanding localization properties of the steady state and dissipation-induced transition, as will shown in the following sections.”
	
	Based on the proposal for realizing quasi-local dissipation operators in 1D disordered or quasi-disordered systems \cite{WYC_PRL,OpenDisorder13}, here the dissipation-driven 3D Anderson model can in principle be experimentally realized in the context of engineered quantum systems. Recent advancements in the engineering of quantum jump operators in open quantum systems, particularly those involving optical lattices and engineered dissipation, have significantly advanced the exploration of dissipation-driven phenomena \cite{Exp_new1,Exp_new2}.
		In particular, the cold atom platform offers a promising experimental realization of the dissipation-driven Anderson model \cite{Anderson_AMO0,Anderson_AMO1,Anderson_AMO2,Anderson_AMO3,Anderson_AMO4,Anderson_AMO5,Anderson_AMO6}. Cold atom experiments, especially those utilizing optical lattices, provide a highly controllable environment where key parameters such as disorder strength, dimensionality, and interactions can be precisely tuned \cite{simulation,simulation1,simulation2}. These methods effectively replicate the dissipation mechanisms described in the 3D Anderson model, offering a powerful platform to explore the delocalized-to-localized transition in the presence of dissipation.

	\section{Dissipation induced transition from delocalization to Localization}
	\label{dissipation_localization}
\begin{figure}[!t]
		\includegraphics[width=8.5cm]{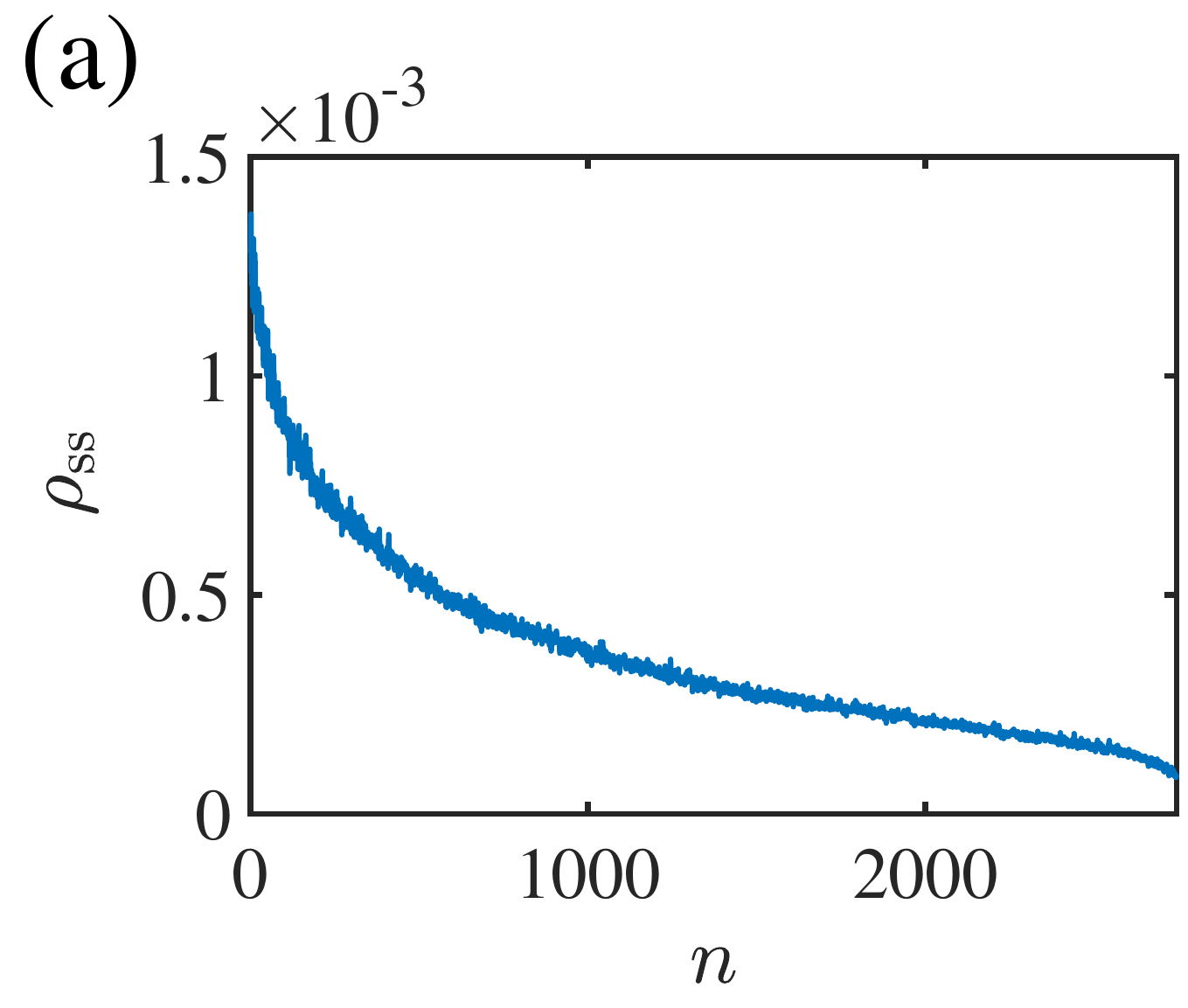}
		\includegraphics[width=8.5cm]{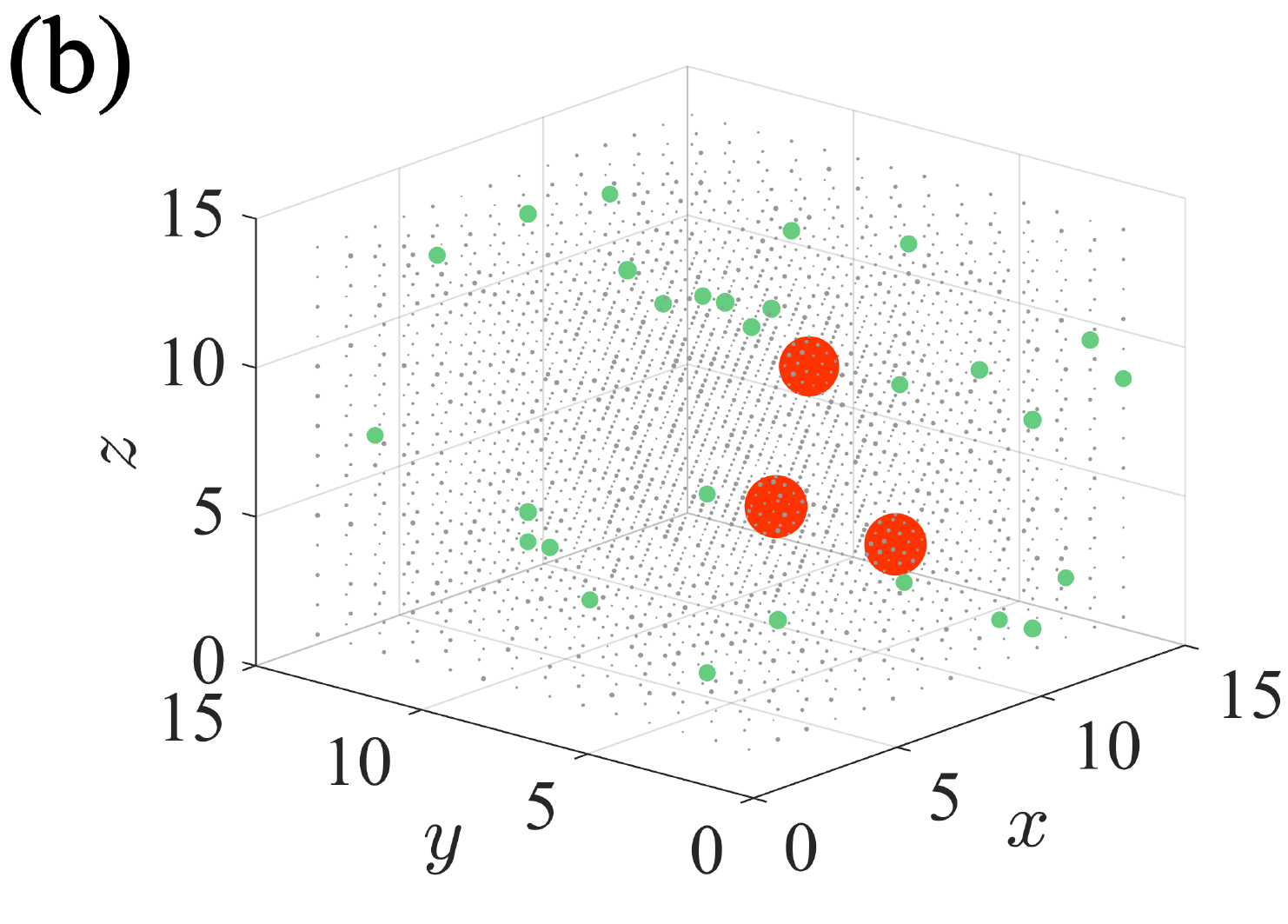}
		\caption{ (a) Diagonal elements of the density matrix for steady state with the dissipative phases $\alpha=0 ,\beta=0 ,\gamma=0$ and $l=1$ in the eigenbasis of Hamiltonian. Here we take $10$ disorder average. The non-diagonal elements (not drawn) are negligible compared with diagonal elements. 
			(b) The probability density of a particle appearing on each lattice site for steady-state without disorder average. Gray blue spheres display the density $|\psi(\boldsymbol{r})|^2 < 0.0008\approx2.2\bar{n}$ (the size of spheres is proportional to the probability density), green spheres show $0.0008<|\psi(\boldsymbol{r})|^2 \leq 0.001 (2.75\bar{n})$ (the size of spheres is proportional to the probability density), and large red dots correspond to $|\psi(\boldsymbol{r})|^2>0.001$ (the size of spheres is proportional to the probability density). Here $\bar{n}=1/L^3$ represents the mean value of probability density and the system parameters are taken as $L=14$, $W=13$ and $\Gamma= 0.1$.}
		\label{Fig2}
	\end{figure}

	In this section, we will see that the dissipation operators introduced in  Eq. \eqref{O_j} can induce a transition from delocalization to localization where the steady state primarily consists of localized states. To show this, we begin by examining the dissipation operators in Eq. \eqref{O_j} for $l_x=l_y=l_z=1$ and first adjust relative phases as $\alpha = \beta = \gamma=0$. In this case, the dissipation operator in each channel reduces to the form $O_{j}^{(m)} = ( c_{\boldsymbol{j}}^\dagger+c_{\boldsymbol{j} + \boldsymbol{1}}^\dagger ) 
	( c_{\boldsymbol{j}} - c_{\boldsymbol{j} + \boldsymbol{1}})$, which drives an anti-symmetric out-of-phase mode to a symmetric 
	in-phase mode. Then we compute the steady-state distribution of density matrix in this kind of dissipation by diagonalizing the corresponding Liouvillian superoperator. The related results are shown in the Fig.\ref{Fig2}. Fig.\ref{Fig2}(a)
	plots the density matrix's steady state distribution on eigenstates of the system Hamiltonian for $W<W_c$. One can see that the steady-state density matrix is 
	mainly located in the low-energy region, composed of localized states. 
	The localization property of the steady-state can be visualized by the density distribution in real space as plotted in Fig.\ref{Fig2}(b). The density distribution on distinct lattice sites differs by one order of magnitude and mainly distributes on a few lattice sites, displaying the localization signature of the steady-state.  
	To reveal why the steady state predominantly concentrates on low-energy localized eigenstates for $\alpha=0,\beta=0,\gamma=0$, we investigate relative phases of neighboring lattice sites for eigenstates of the system Hamiltonian. Fig.\ref{fig_IPR_Phase}(b) plots the proportion of in-phase site pairs $P^{\text{in}}_{n,1}$ which is monotone decreasing as the energy growth from which we can see clearly that the localized states in the low-energy side of the spectrum have more in-phase site pairs. In contrast, the delocalized states in the middle of the
	spectrum tend to obtain more (less) out-of-phase (in-phase) site pairs. Since dissipation operators tend to drive the system to symmetric in-phase modes, the distribution $P^{\text{in}}_{n,1}$ illustrates the physical origin of the steady state distribution favoring those eigenstates with lower energies.

	\begin{figure}[!t]
		\includegraphics[width=8.5cm]{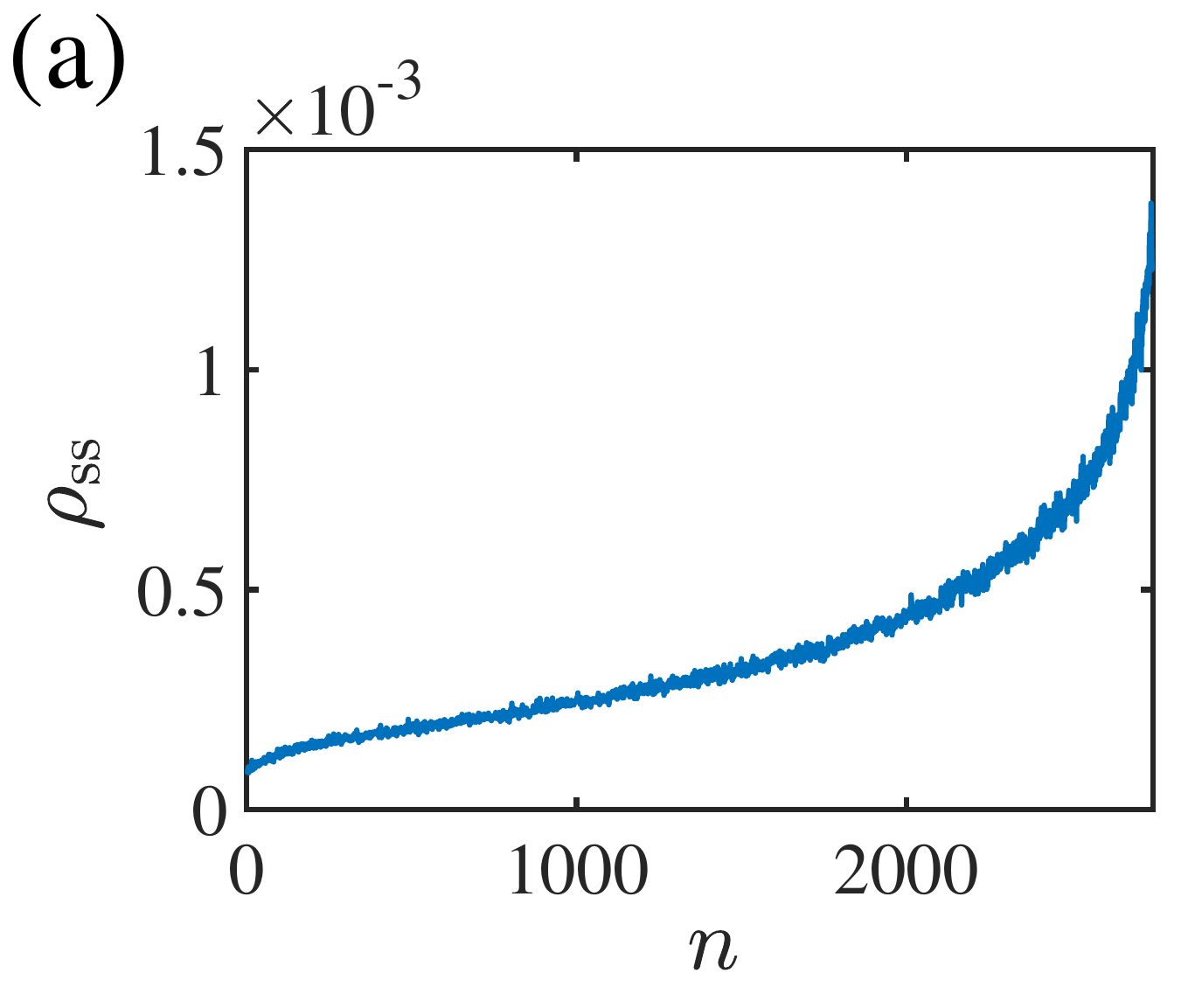}
		\includegraphics[width=8.5cm]{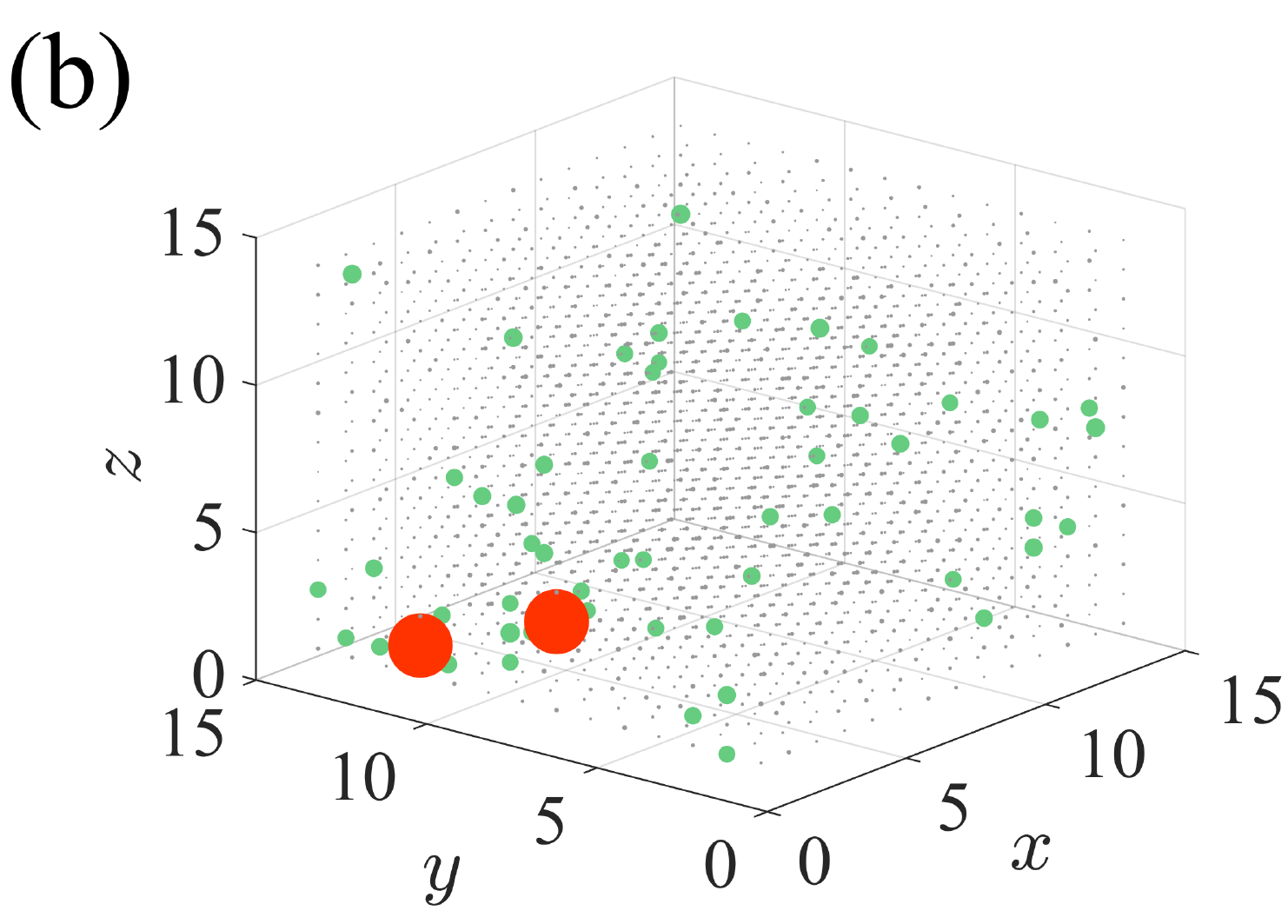}
		\caption{ (a) Diagonal elements of the density matrix for steady state with the dissipative phases $\alpha=\pi ,\beta=\pi ,\gamma=\pi$ and $l=1$ in the eigenbasis of Hamiltonian. Here we take $10$ disorder average. The non-diagonal elements (not drawn) are negligible compared with diagonal elements. 
			(b) The probability density of a particle appearing on each lattice site for steady state without disorder average. Grey spheres display the density $|\psi(\boldsymbol{r})|^2 < 0.0008\approx2.2\bar{n}$ (the size of spheres is proportional to the probability density), green spheres show $0.0008<|\psi(\boldsymbol{r})|^2 \leq 0.001 (2.75\bar{n})$ (the size of spheres is proportional to the probability density), and large red dots correspond to $|\psi(\boldsymbol{r})|^2>0.001$ (the size of spheres is proportional to the probability density). Here $\bar{n}=1/L^3$ represents the mean value of probability density and the system parameters are taken as $L=14$, $W=13$ and $\Gamma= 0.1$. }
		\label{Fig3}
	\end{figure}
	
	We then study the situation of relative phases as $\alpha = \beta = \gamma=\pi$ in dissipation operators $l_x=l_y=l_z=1$ where the dissipation operator in each channel reduces to the form $O_{j}^{(m)} = ( c_{\boldsymbol{j}}^\dagger-c_{\boldsymbol{j} + \boldsymbol{1}}^\dagger ) 
	( c_{\boldsymbol{j}}+ c_{\boldsymbol{j} + \boldsymbol{1}})$. In this case, the dissipation drives the system from an in-phase mode to an out-of-phase mode.
	The corresponding steady-state distribution of the density matrix on the eigenbasis of the Hamiltonian is shown in Fig.\ref{Fig3}. From Fig.\ref{Fig3}(a), one can see that the steady state of the density matrix is mainly located in the high-energy region composed of localized states. Similar to discussions about the case of  $\alpha=0,\beta=0,\gamma=0$, the localization property can be visualized in real space as plotted in Fig.\ref{Fig3}(b) where The density distribution on distinct lattice sites differs by an order of magnitude and mainly distributes on a few lattice sites.  
	To understand the mechanism of steady state distribution, we also employ the proportion of in-phase pairs $P^{\text{in}}_{n,1}$ for eigenstates as discussed above. As stated above, the states in low-energy
	side of the spectrum tend to obtain more in-phase
	site pairs. In contrast, those states on the high-energy side of the spectrum tend to have more out-of-phase site pairs. This is because the relative phase of each eigenstate for any site pair is either $0$ (in phase) or $\pi$ (out of phase). Thus, the steady state predominantly concentrates on high-energy regions consisting of localized states for relative phases as $\alpha = \beta = \gamma=\pi$.
	
\begin{figure}[!t]
		\includegraphics[width=8.5cm]{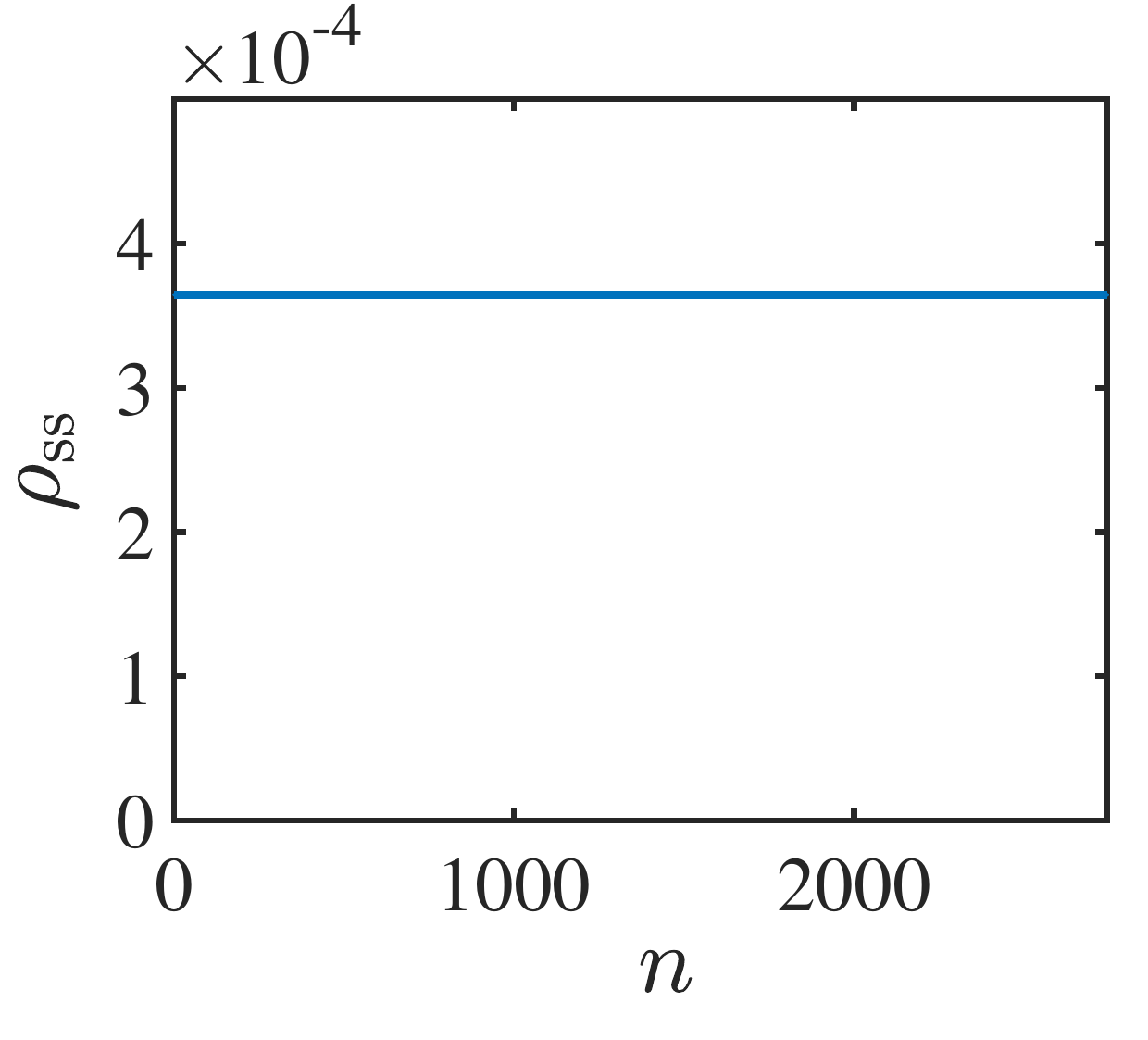}
		\caption{Diagonal elements of the density matrix for steady state with the dissipative phases $\alpha=\pi/2 ,\beta=\pi/2,\gamma=\pi/2$ and $l=1$ in the eigenbasis of Hamiltonian. All diagonal elements are same and non-diagonal elements are exactly zero (not drawn) which means the density matrix is proportional to the identity matrix. The parameters are taken as $L=14$, $W=13$ and $\Gamma= 0.5$.}
		\label{Fig4}
	\end{figure}

	Therefore, if the initial state is prepared on a delocalized state, the system can be driven to a steady state predominantly consisting of localized states, meaning that a localization transition is implemented by using a tailored dissipation $\alpha = \beta = \gamma=0$ or $\alpha = \beta = \gamma=\pi$ for $l=1$. On the other hand, the Anderson localization is maintained if we prepare a localized initial state under the above dissipation.
	It is important to note that the above consequence is highly non-trivial because the Anderson localization is generally fragile under dissipation. For instance, if the system couples with reservoirs via local density $O_{j}=n_j$  describing pure dephasing at lattice site $j$, which would destroy the Anderson localization. 
	The fragility can also be seen clearly if one simply chooses relative phases as $\alpha = \beta = \gamma=\pi/2$ in Eq.\eqref{O_j}, and  each dissipation operator is hermitian $O_{j}^{(m)} = ( c_{\boldsymbol{j}}^\dagger+ic_{\boldsymbol{j} + \boldsymbol{1}}^\dagger) 
	( c_{\boldsymbol{j}} - ic_{\boldsymbol{j} + \boldsymbol{1}})=O_{j}^{(m)\dagger}$. Since all dissipation operators are hermitian, the system will relax to the maximally mixed state where the density matrix (as shown in Fig.\ref{Fig4}) is proportional to identity as its steady-state \cite{Longhi2023}, leading to the destruction of the Anderson localization. 
	
	\section{Dissipation induced transition from localization to delocalization}
	\label{dissipation_extension}
\begin{figure*}[!ht]
		\includegraphics[width=0.98\linewidth]{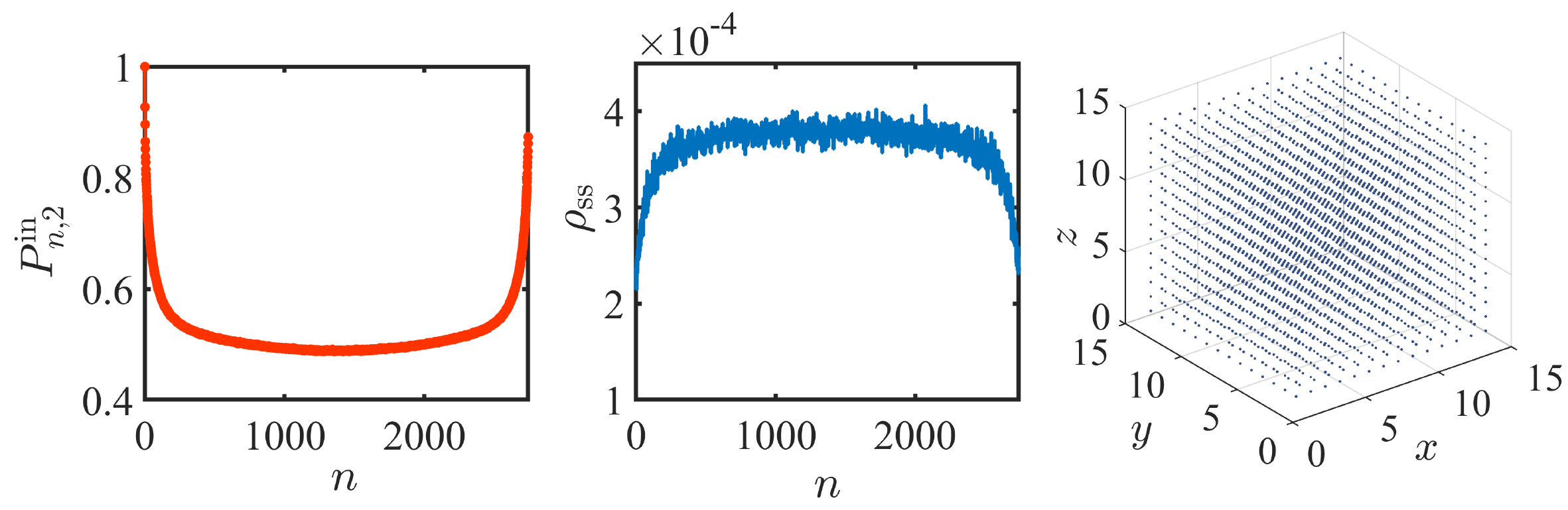}
		\caption{Left panel: The proportion of in-phase site pairs $P_{n,2}^{\text{in}}$ for each eigenstate with the disorder average of $100$ times. Middle panel: Diagonal elements of the steady-state density matrix for the dissipative phases $\alpha=\pi ,\beta=\pi ,\gamma=\pi$ and $l=2$ in the eigenbasis of Hamiltonian with the disorder average of $10$ times. The non-diagonal elements (not drawn) are negligible compared with diagonal elements.  Right panel: The probability distribution of a particle  appearing on each lattice point for the steady-state. The density is distributed within the interval $\big[0.9\bar{n},1.1\bar{n}\big]$ with mean value $\bar{n}=1/L^3$ showing nearly uniform distribution. Here the system parameters are chosen as $L=14$, $W=13$ and $\Gamma=1.0$. }
		\label{Fig5}
	\end{figure*}

	As described in the previous section, it has shown that the dissipation for $l=1$ can drive the system into a steady state mainly composed of localized states. It naturally raises a question of how to make the system relax to a steady state exhibiting delocalization instead of localization property. In the previous work \cite{WYC_PRL}, a dissipation-induced transition from localization to delocalization is realized by setting $l=2$ where the steady state primarily concentrates on the middle of the spectrum with delocalized eigenstates. 
	Inspired by the work, we further explore the effect of dissipation operators in Eq.\eqref{O_j} with $l=2$. To determine the relative phases in the dissipation operators, we first investigate the proportion of in-phase lattice site pairs, i.e., $P^{\text{in}}_{n,2}$. From the Fig.\ref{Fig5} (left panel), we can see that $P^{\text{in}}_{n,2}$ exhibits a shallow U-shaped pattern where the localized states on both
	sides of the spectrum exhibit more in-phase pairs, while the delocalized states in the middle of spectrum have more (less) out-of-phase (in-phase) site pairs which obviously differ from $P^{\text{in}}_{n,l}$ in the case of $l=1$ (see Fig.\ref{fig_IPR_Phase}(b)). This consequence prompts us to investigate and select the dissipative phases as $\alpha=\beta=\gamma=\pi$ and survey the steady-state distribution of the density matrix. As shown in Fig. \ref{Fig5} (middle panel), the system is anticipated to reach a steady state predominantly composed of those states associated with out-of-phase site pairs, thereby primarily favoring the dominance of delocalized eigenstates in middle energy region. 
	
	Since the steady state here is predominately composed of delocalized eigenstates, it should be anticipated to exhibit delocalization property. To show this, we visualize the density distribution of the steady-state in real space as plotted in Fig. \ref{Fig5} (right panel), where the distribution uniformly spread over the whole lattice showing significant difference from the distribution in Fig. \ref{Fig2}(b) and Fig. \ref{Fig3}(b).
	Therefore, by choosing the dissipation
	phase $\alpha=\beta=\gamma=\pi$ and the distance $l=2$, we can realize the transition from localization to delocalization. 
	
	\section{Conclusion and outlook} 
	\label{summarize}
	
	We have conducted a comprehensive investigation into the effects of dissipation on the 3D Anderson model, which features mobility edges that delineate delocalized from localized states. Through the analysis of steady-state density matrix distributions, we discovered that dissipation can steer the system towards specific states predominantly composed of either localized or delocalized states, irrespective of the initial conditions. The characteristics of these steady states are intricately connected to the dissipative phases within the dissipation operators. Notably, when the dissipation operators couple nearest neighbor site pairs ($l=1$) with phases $\alpha=\beta=\gamma=0$ or $\alpha=\beta=\gamma=\pi$, the system tends to adopt a specific localized state. Conversely, dissipation operators that couple next nearest neighbor site pairs ($l=2$) with phases $\alpha=\beta=\gamma=\pi$ propel the system towards delocalized states. This implies that by carefully adjusting the parameters, dissipation can induce transitions between localized and delocalized phases. Consequently, dissipation emerges as a powerful tool for facilitating transitions between localized and delocalized states and for manipulating transport properties.
	
	Our research substantiates that dissipation can guide a 3D disordered system into a unique steady state, predominantly marked by either delocalized or localized states, rather than merely disrupting them. These findings offer novel insights into the manipulation of quantum systems through dissipation. Drawing inspiration from the precise control over dissipation in quantum systems, as seen in condensed matter and cold atom experiments, our results hold promising implications for quantum technology and simulation. Beyond single-particle disordered systems, our study also opens avenues for exploring many-body systems that exhibit non-thermal properties. Our findings suggest potential strategies for managing transitions between thermalized states and many-body localized states, or other non-thermal states that defy the eigenstate thermalization hypothesis (ETH)  \cite{Diss_scar}.
	
    Recent experimental advancements have demonstrated the feasibility of realizing dissipative processes in quantum systems, particularly in cold-atom setups. For instance, experiments involving engineered dissipation in optical lattices have successfully demonstrated the control of quantum states through dissipative mechanisms \cite{simulation1,simulation2}. Specifically, experiments have been realized using spin-dependent lattices and auxiliary sites \cite{Jenkins2022,Chen2022}, where the dissipative coupling between different sites with distant $l=1$ and $l = 2$ can be achieved through carefully designed laser configurations \cite{WYC_PRL}. These experiments provide a strong foundation for the experimental realization of our proposed dissipative mechanisms and highlight the potential for observing dissipation-induced delocalized-localized transitions in real quantum systems. 
	
	Moreover, the dissipation operators employed in this study leverage phase distribution, offering a fresh perspective for investigating other experimentally viable dissipative mechanisms to achieve specific quantum states. This approach not only enhances our understanding of quantum state manipulation but also paves the way for innovative applications in quantum computing and materials science.
	
	\section*{Acknowledgements}
	The work is supported by the National Key R\&D Program of China under Grant No.2022YFA1405800 and National Natural Science Foundation of China (Grant No. 12304290). LP also acknowledges support from the Fundamental Research Funds for the Central Universities. \\

	\subsection{Appendix}
	\label{appendix:A}
	 In this appendix, we detailed derivation of the the Liouvillian superoperator formula using the Choi-Jamiołkowski isomorphism. We begin with the Lindblad master equation 
\begin{align}
	\frac{d\rho(t)}{dt} = -i[H_S, \rho(t)] + \sum_{j} \sum_{m=1}^M \Gamma_j^{(m)} &\Big( O_j^{(m)} \rho O_j^{(m)\dagger}\nonumber \\
	&- \frac{1}{2} \{ O_j^{(m)\dagger} O_j^{(m)}, \rho \} \Big).\nonumber 
\end{align}
	Using the Choi-Jamiołkowski isomorphism, the density matrix \(\rho\) is vectorized as \(|\rho\rangle = \mathrm{vec}(\rho)\), where \(\mathrm{vec}(\rho)\) stacks the columns of \(\rho\) into a single column vector. The vectorization operation follows 
	\begin{align} \mathrm{vec}(A \rho B) &= (B^{\mathrm{T}} \otimes A) \, \mathrm{vec}(\rho), \nonumber\\ \mathrm{vec}([A, \rho]) &= (A \otimes I - I \otimes A^{\mathrm{T}}) \, \mathrm{vec}(\rho), \nonumber\\ \mathrm{vec}(\{A, \rho\}) &= (A \otimes I + I \otimes A^{\mathrm{T}}) \, \mathrm{vec}(\rho). 
	\end{align}
	In this way, the unitary part of the Lindblad equation \(-i[H_S, \rho]\) is transformed as 
	\begin{align}
	\mathrm{vec}(-i[H_S, \rho]) = -i(H_S \otimes I - I \otimes H_S^{\mathrm{T}}) \, |\rho\rangle,
	\end{align}
	and the dissipative part of the Lindblad equation is transformed as
	\begin{align}
		\mathrm{vec}(O_j^{(m)} \rho O_j^{(m)\dagger}) &= (O_j^{(m)} \otimes O_j^{(m)*}) \, |\rho\rangle, \nonumber \\
		\mathrm{vec}(\{O_j^{(m)\dagger} O_j^{(m)}, \rho\}) &= (O_j^{(m)\dagger} O_j^{(m)} \otimes I + I \otimes (O_j^{(m)\dagger} O_j^{(m)})^{\mathrm{T}}) \, |\rho\rangle.
	\end{align}
	Therefore, the Liouvillian superoperator acts on the vectorized density matrix as:$ \frac{d|\rho\rangle}{dt} = \mathscr{L} |\rho\rangle$ where the total Liouvillian superoperator is obtained by combining the unitary and dissipative contributions 
	\begin{widetext}
    \begin{align}
		\mathscr{L} = -i(H_S \otimes I - I \otimes H_S^{\mathrm{T}}) + \sum_{j} \sum_{m=1}^M\Gamma_j^{(m)} \left[ 2 O_j^{(m)} \otimes O_j^{(m)*}- O_j^{(m)\dagger} O_j^{(m)} \otimes I - I \otimes \Big(O_j^{(m)\dagger} O_j^{(m)}\Big)^{\mathrm{T}} \right]. \nonumber
		\end{align} 
	\end{widetext}

\end{document}